\documentclass[12pt]{article}
\usepackage{graphicx}
\usepackage{psfrag}

\def\beq{\begin{eqnarray}}
\def\eeq{\end{eqnarray}}
\def\M{M_{(5)}}

\def\mpl{M_{\rm Pl}}

\def\R{{\tilde R}}
\def\g{\tilde g}

\begin{document}

\begin{flushright}
NYU-TH/01/04/03 \\
TPI-MINN-01/17\\
UMN-TH-2003 \\
astro-ph/0105068\\
\end{flushright}

\vskip 1cm
\begin{center}
{\Large \bf
Accelerated Universe from Gravity Leaking to Extra Dimensions}\\
\vskip 2cm
{C\'edric Deffayet, Gia Dvali, Gregory Gabadadze$^*$}\footnote
{e-mail cjd2@physics.nyu.edu\\

~~e-mail gd23@nyu.edu\\

~~e-mail gabadadz@physics.umn.edu}\\ 
\vskip 1cm 
{\it Department of Physics, New York University, New York, NY 10003\\
$^*$Theoretical Physics Institute, University of Minnesota, Minneapolis, 
MN 55455}\\
\end{center}

\vspace{0.9cm}
\begin{center}
{\bf Abstract}
\end{center}
We discuss the idea that the accelerated Universe could be the 
result of the gravitational leakage into extra dimensions on Hubble 
distances  rather than the consequence of non-zero cosmological 
constant.

\vspace{0.1in}

\newpage

\section{Introduction}

A number of recent observations suggest that the Universe is accelerating at
large scales \cite {cc} (see also \cite{CMB1,CMB2}). 
This may be regarded as an  evidence for non-zero but
very small cosmological constant. However, before adopting such a 
conclusion it is desirable to explore alternative possibilities
motivated by  particle physics considerations. In this respect the models
which predict modification of gravity at large distances are particularly
interesting.  In the present paper we focus on the  5-dimensional
brane-world model with infinite-volume extra dimension 
which can predict such a modification at cosmological 
distances \cite {DGP,DG}. 
In this model the ordinary particles are
localized on a  3-dimensional surface (3-brane) embedded in infinite
volume extra space to which gravity can spread. Despite the presence of an
infinite volume flat extra space, the observer on the brane measures
four-dimensional Newtonian gravity at distances shorter than a certain
crossover scale $r_c$ which can be of astronomical size \cite {DGP,DG}.
This phenomenon is due to a
four-dimensional Ricci scalar term which is induced on the brane 
\cite {DGP,DG}.
The whole dynamics of gravity is governed by competition between this term
and an ordinary five-dimensional Einstein-Hilbert action.
At short distances the four-dimensional term dominates and ensures that
gravity looks four-dimensional. At larger distances, however, the
five-dimensional term takes over and gravity spreads into extra
dimensions. As a result, the force law becomes 5-dimensional one. 
Thus, gravity gets
weaker at cosmic distances.
It is natural that such a dramatic modification should affect the
cosmological expansion of the Universe. In the present work we 
will focus on the explicit
cosmological solution found in \cite{Cedric}. This solution
describes the Universe which is accelerated beyond the cross-over
scale. The acceleration takes place despite the fact that there 
is no cosmological constant on the brane. 
Instead, the bulk gravity sees its own curvature term on the brane as
cosmological constant and accelerates the Universe.

In the present paper we shall review this phenomenon in the light of
recent astrophysical observations \cite {cc,CMB1,CMB2}
and confront this model with the conventional cosmological constant
scenario. 
We shall show that the present  scenario cannot be
mimicked by ordinary 4D gravity with arbitrary high-derivative terms.
Therefore, this is intrinsically high-dimensional phenomenon. 
Finally, we argue that such
scenarios might avoid difficulties of reconciliation of String Theory 
with the observation of the accelerated Universe.
This is possible due to the fact that the bulk metric in the theory 
is Minkowskian. Moreover, due to the leakage of gravity   
into  extra space there is no infinite 
future horizon for 4D observers.
 
Before we proceed we would like to note that
other interesting cosmological solutions in this type of models were 
first studied in Ref. \cite {Csaki}, however, those solutions do
not describe accelerated Universe and will not be discussed here.

\section{The Framework}

The model we will be considering was introduced in Ref. \cite {DGP}.
We start with a $D=(4+1)$ dimensional theory. Let us suppose
there is  a 3-brane
embedded in $5$-dimensional space-time\footnote {For simplicity
we ignore brane fluctuations in which case the induced metric 
on the brane takes a simple form given below in (\ref {4Dg}).}. 

Four coordinates of our world are  $x_\mu,~\mu=0,1,2,3$;
the extra coordinate will be denoted by  $y$.
Capital letters and subscripts will be used for
5D quantities  ($A,B,C=0,1,2,3,5$);
the metric convention is mostly positive.

Following Ref. \cite {DGP,DG} let us consider the action:
\beq
S~=~ {\M^3\over 2}~\int d^5X ~\sqrt {|\g|}~{\R}~+~
{\mpl^2\over 2}~\int d^4x ~\sqrt {|g|}~{R}~,
\label{1}
\eeq
where $\M$ denotes the 5D Planck mass, and $\mpl$
is the 4D Planck mass; as they stand in (\ref {1}) $\M$ and $\mpl$
are independent parameters
(in general they could  be related). 
$\g_{AB}(X) \equiv \g_{AB}(x,y)$ denotes
a 5D metric for which the 5D Ricci scalar is
${\R}$. The brane is located at $y=0$.
The induced metric on the  brane is denoted by
\beq
g_{\mu\nu}(x)~\equiv~\g_{\mu\nu} (x,~y=0)~.
\label{4Dg}
\eeq
The 4D Ricci scalar for $g_{\mu\nu}(x)$ is $R=R(x)$.
The Standard Model (SM)
fields are confined to the brane. Note that the SM  cutoff
should not coincide in general with $\M$ and,  in fact, 
is assumed to be much higher in our case. For simplicity   
we suppress the Lagrangian of SM fields.
The braneworld origin of the action (\ref {1}) and parameters  $\M$, $\mpl$ 
were discussed in details in Refs. \cite {DGP,DG,DGKN}. 

Let us first study the non-relativistic potential between 
two sources confined to the brane.
For a time being we drop the tensorial structure in
the gravitational equations and discuss the distance dependence of the
potential. We comment on the tensorial structure at the end of this 
section.

The static gravitational potential between the  
sources in the 4-dimensional
world-volume of the brane is determined as:
\beq
V(r)~=~\int G_R\left (t,{\overrightarrow x},y=0; 0,0,0\right
) dt~,
\label{pot}
\eeq
where $r\equiv\sqrt{x_1^2+x_2^2+x_3^2}$ and 
$G_R\left (t,{\overrightarrow x},y=0; 0,0,0\right )$
is the retarded Green's function (see below).
Let us turn to
Fourier-transformed quantities with respect to
the world-volume four-coordinates $x_\mu$:
\beq
G_R(x,y; 0,0)~\equiv~\int ~ {d^4p\over (2\pi)^4}~e^{ipx} ~{\tilde G}_R(p,y)~.
\label{Fourie}
\eeq
In  Euclidean momentum space the equation for the Green's function
takes the form:
\beq
\left (~ \M^3(p^2-\partial_y^2)~+~\mpl^2~ p^2 ~\delta(y) ~\right )~
{\tilde G}_R(p,y)~=
~\delta(y)~.
\label{mom}
\eeq
Here $p^2$ denotes the square of an Euclidean four-momentum
$p^2\equiv p_4^2+p_1^2+p_2^2+p_3^2$.
The solution with appropriate boundary conditions
takes the form:
\beq
{\tilde G}_R(p,y)~=~{1\over \mpl^2p^2~+~2\M^3p}~ {\rm exp} (-p|y|)~,
\label{sol1}
\eeq
where $p\equiv\sqrt {p^2}=\sqrt{p_4^2+p_1^2+p_2^2+p_3^2}$.
Using this expression and Eq. (\ref {pot}) one finds the following
(properly normalized)
formula for the  potential
\beq
V(r)~=~-{1\over 8\pi^2 \mpl^2}~{1 \over r}~\left \{ {\rm sin}
\left ( {r\over r_c} \right ) ~{\rm Ci} \left ( {r\over r_c} \right )
~+~{1\over 2}  {\rm cos}
\left ( {r\over r_c} \right ) \left
[\pi~-~2 ~ {\rm Si} \left ( {r\over r_c} \right ) \right ]   \right \}~,
\label{V}
\eeq
where
$ {\rm Ci}(z) \equiv \gamma +{\rm ln}(z) +\int_0^z ({\rm cos}(t) -1)dt/t$,
$ {\rm Si}(z)\equiv \int_0^z {\rm sin}(t)dt/t$,
$\gamma\simeq 0.577$  is  the Euler-Mascheroni
constant, and the distance scale $r_c$ is defined as follows:
\beq
r_c~\equiv ~{\mpl^2\over 2 \M^3}~.
\label{r0}
\eeq
In our model we choose $r_c$ to be of the order of
the present Hubble size,  which is equivalent 
 to the choice $\M \sim 10-100$ MeV. We will discuss 
 phenomenological compatibility
  of such a low quantum gravity scale in section 6. 
It is useful to study  the short distance and long distance
behavior of this expression.

At short distances when $r<<r_c$ we find:
\beq
V(r)~\simeq~-{1\over 8\pi^2 \mpl^2}~{1 \over r}~\left \{
{\pi\over 2} +\left [-1+\gamma+{\rm ln}\left ( {r\over r_c} \right )
\right ]\left ( {r\over r_c} \right )~+~{\cal O}(r^2)
\right \}~.
\label{short}
\eeq
Therefore, at short distances the potential
has the correct 4D Newtonian $1/r$ scaling. 
This is subsequently modified
by the logarithmic {\it repulsion} term in (\ref {short}).

Let us turn now to the large distance behavior. Using (\ref {V})
we obtain for $r>>r_c$:
\beq
V(r)~\simeq~-{1\over 8\pi^2 \mpl^2}~{1 \over r}~\left \{
{r_c\over r}~+~{\cal O} \left ( {1\over r^2} \right )
\right \}~.
\label{long}
\eeq
Thus, the long distance potential
scales as $1/r^2$ in accordance with laws of 5D theory.

We would like to emphasize that the behavior (\ref {sol1})
is intrinsically higher-dimensional and 
is very hard to reproduced in conventional four-dimensional 
field theory. Indeed, the would be four-dimensional 
inverse propagator should contain the term $\sqrt{p^2}$.
In the position space this would correspond in the Lagrangian to  
the following pseudodifferential operator 
\beq
{\hat {\cal O}}~=~-~\partial_\mu^2~+~{\sqrt{-\partial_\mu^2}\over r_c}~.
\label{root}
\eeq
We are not aware 
of a consistent four-dimensional quantum field theory 
with a finite number of physical bosons which would lead to such 
an effective action. 

Finally we would like to comment on the 
tensorial  structure of the graviton propagator in the 
present model. In flat space this structure is similar to that of
a massive 4D graviton \cite {DGP}. 
This points to the van Dam-Veltman-Zakharov 
(vDVZ) discontinuity \cite {Veltman,Zakharov}. 
However, this problem can in general be resolved by at least 
two methods. In the present context one has to use the results
of \cite {Vainshtein} where it was argued that 
the vDVZ discontinuity which emerges
in the lowest perturbative approximation is in fact absent in the 
full nonperturbative theory. The application of the similar arguments 
to  our model leads to the result which is continuous in $1/r_c$. 
This will be discussed in details elsewhere \cite {disc}.
Thus, the vDVZ problem is an artifact of using 
the lowest perturbative approximation\footnote{
Note that the continuity in the graviton mass in (A)dS backgrounds was
demonstrated recently in Refs. \cite {Kogan,Massimo}. We should emphasize 
that we are discussing the continuity in the classical 
4D gravitational interactions on the brane.
There is certainly the discontinuity in the full theory in a sense that 
there are extra degrees of freedom in the model. These latter
can manifest themselves at quantum level in loop diagrams 
\cite {Duff}.}. 

In general, the simplest possibility to 
deal with the vDVZ problem, as was suggested in Ref.
\cite {DGKN}, is to compactify the extra space
at scales bigger than the Hubble size with 
$r_c$ being even bigger, 
but we do not consider this possibility here.

\section{Cosmological Solutions}

Below we  will mainly be interested in  
the geometry of the 4D brane-world.  For the completeness of the presentation
let us first recall the full  5D metric of the cosmological solution. 
The 5D line element  is taken in the following form: 
\beq
ds^2~=~-N^2(t,y)~dt^2~+~A^2(t,y)~\gamma_{ij}~dx^idx^j~+~B^2(t,y)~dy^2~,
\label{5Dint}
\eeq
where $\gamma_{ij}$ is the metric of a 3 dimensional maximally 
symmetric Euclidean 
space, and the metric coefficients read \cite {Cedric}
\begin{eqnarray}
N(t,y)~&=&~1~+~\epsilon~|y|~{\ddot a}~({\dot a}^2~+~k)^{-1/2}~, \nonumber \\
A(t,y)~&=&~a~+~\epsilon~|y|~({\dot a}^2~+~k)^{1/2}~, \nonumber \\
B(t,y)~&=&~1~,
\label{nab}
\end{eqnarray}
where $a(t)$ is 4D scale factor and $\epsilon =\pm 1$.
Knowing the braneworld intrinsic geometry is all what matters
as far as 4D observers are concerned.
This geometry is given in the above solution. 
Taking the $y=0$ value of the metric we obtain  the 
usual 4D Friedmann-Lema\^{\i}tre-Robertson-Walker (FLRW) 
form (enabling to interpret $t$ as the cosmic time on the braneworld)
\begin{eqnarray} \label{FLRW}
ds^2~ &=&~ -dt^2~ + ~a^2(t)~ dx^i~ dx^j ~\gamma_{ij},\\
&=&  ~-dt^2~ + ~a^2(t)~\left(dr^2 + S^2_k(r)d\psi^2 \right),
\end{eqnarray}
where $d\psi^2$ is an  angular line element, $k=-1,0,1$ parametrizes
 the brane world  spatial curvature, and $S_k$ is given by 
\begin{eqnarray}
S_k(r) = \left\{\begin{array}{ll} \mbox{sin } r & (k=1) \\
\mbox{sinh } r &(k=-1) \\ r &(k=0) \end{array} \right \}
\end{eqnarray}
In the present case,
the dynamics is generically different from the 
usual FLRW cosmology, as shown in  \cite{Cedric}.
The standard first Friedmann equation is replaced in our model 
by 
\begin{equation} \label{fried}
H^2 + \frac{k}{a^2} = \left(\sqrt{\frac{\rho}{3 {\mpl^2}}  +
 \frac{1}
{4 r_c^2}} +\epsilon \frac{1}{2 r_c}\right)^2, 
\end{equation}
where $\rho$ is the total cosmic fluid energy density.
We have in addition the usual equation of conservation for 
the energy-momentum tensor of the cosmic fluid given by 
\beq \label{cons}
\dot{\rho} + 3H (p + \rho) = 0~.
\eeq
Equations (\ref{fried}) and (\ref{cons}) are sufficient 
to derive the cosmology of our model. In particular using 
these relations
one can obtain a second Friedmann equation as in 
standard cosmology.

Equation (\ref{fried}) with 
$\epsilon =1$ and $\rho =0$ has an interesting self-inflationary solution
with a Hubble parameter given by the 
inverse of the crossover scale $r_c$.
This can be easily understood looking back at the action  (\ref{1}) 
where it is apparent that the intrinsic curvature term on the brane 
appears as a source for the bulk gravity, so that with appropriate 
initial conditions, this term can cause  an expansion of the 
brane world without the need of matter or cosmological constant 
on the brane. This self-inflationary solution is the key ingredient for 
our model to produce late time accelerated expansion\footnote{
Note that the nonzero 4D Ricci scalar on the brane makes a seemingly 
 negative contribution to the 
brane tension \cite{Zurab,Cedric}. In this case, we consider a non fluctuating brane which is placed at the 
${\bf R}/{\bf Z}_2$ orbifold fixed point.}. Before discussing in detail this issue let us first  
compare our cosmology with the the standard one. 

We first note that the standard cosmological evolution is recovered 
from (\ref{fried}) whenever $\rho / {\mpl^2}$ is large  
compared to  $1/ r_c^2$, so that the early time cosmology of 
our model is analogous to standard cosmology. In this early phase
equation (\ref{fried}) reduces, at leading order,
to the standard 4D Friedmann equation given by 
\begin{equation} \label{stanfried}
H^2+ \frac{k}{a^2}~ = \frac{\rho}{3 {\mpl }^2}~.\end{equation}

The late time behavior is however generically different, as was 
shown in \cite{Cedric}: when the energy density decreases
and crosses the threshold $\mpl^2/r_c^2$,
 one either has a transition to a pure $5D$ regime (see e.g. \cite{bdl,bdel}) where the Hubble 
parameter is linear in
 the energy density $\rho$ (this happens for the $\epsilon =-1$ branch of the 
solutions), 
or to the self inflationary solution mentioned above (when $\epsilon =+1$).
This latter is the case we would like to investigate in more detail 
 in the rest of this work and we set $\epsilon =+1$ from now on. 
In terms of the Hubble radius (and for the flat Universe) 
the crossover between the 
two regimes happens when the Hubble radius $H^{-1}$
is of the order of the crossover length-scale between 
4D and 5D gravity,
that is $r_c$. 
If we do not want to spoil the successes of the ordinary cosmology, 
we have thus to assume the $r_c$ is of the order of the present 
Hubble scale $H_0^{-1}$. 

The conservation equation (\ref{cons}) is the same as the standard one, 
so that a given component of the cosmic fluid (non relativistic matter, 
radiation, cosmological constant...) 
will have the same dependence on  the scale factor as in standard cosmology. 
For instance, for a given component, labeled by $\alpha$, which has the 
equation of state $p_\alpha = w_\alpha \rho_\alpha$ (with $w_\alpha$ being a constant) one gets from  (\ref{cons})
$\rho_\alpha  = \rho_\alpha^0 a^{-3(1+ w_{\alpha})}$ 
(with $\rho_\alpha^0$ being a constant). The 
Friedmann equation 
(\ref{fried}) can be rewritten in term of the red-shift 
$1+z \equiv a_0/a$ as follows: 
\begin{equation} \label{H5D}
H^2(z) = H_0^2 \left\{ \Omega_k (1+z)^2 + 
\left( \sqrt{\Omega_{r_c}} + 
\sqrt{ \Omega_{r_c} + \sum_\alpha \Omega_\alpha(1+z)^{3(1+w_\alpha)} } 
\right)^2 \right\},
\end{equation}
where the sum is over all the components of the cosmic fluid.
 In the above equation  $\Omega_\alpha$ is defined as follows: 
\begin{equation} \label{omegafried}
\Omega_\alpha \equiv \frac{\rho^0_\alpha}{3 {\mpl}^2 
H_0^2 a_0^{3(1+w_\alpha)}}~,
\end{equation}
while $\Omega_k$ is given by 
\begin{equation}
 \Omega_k \equiv \frac{-k}{H_0^2 a_0^2}~,
\end{equation}
and $\Omega_{r_c}$ denotes  
\begin{equation}
\Omega_{r_c} \equiv \frac{1}{4 r_c^2 H_0^2}~.
\end{equation}
In the rest of this paper, as far as the cosmology of our model 
is concerned  we will consider a  
non-relativistic matter with density $\Omega_M$
in which case equation (\ref{H5D}) reads\footnote{Notice that we have set the cosmological constant 
on the brane to zero, 
and will do so until the end of this work since we are interested 
here in producing an accelerated Universes without cosmological constant.}.
\begin{equation} \label{HH5D}
H^2(z) = H_0^2 \left\{ \Omega_k (1+z)^2 + 
\left( \sqrt{\Omega_{r_c}} + 
\sqrt{ \Omega_{r_c} +  \Omega_M(1+z)^3 } 
\right)^2 \right\}.
\end{equation}
We can compare this equation with  the conventional 
Friedmann equation:
\begin{equation} \label{HH4D}
H^2(z) = H_0^2 \left\{ \Omega_k (1+z)^2 + 
\Omega_M(1+z)^3 +\Omega_X(1+z)^{3(1+w_X)}  \right\}~.
\end{equation}
Here, in addition to the matter and curvature contributions 
we have included the density of a dark energy component $\Omega_X$ 
with equation of state parameter $w_X$.
When $w_X=-1$, the dark energy acts in the same 
way as a cosmological constant, 
and the corresponding $\Omega_X$ will be denoted 
as $\Omega_\Lambda$ in the following. 
Comparing (\ref{HH5D}) and (\ref{HH4D}) we see that  
$\Omega_{r_c}$ acts similarly (but not identically, as we will see below) 
to a cosmological constant. 

\begin{figure}
\centering
\psfrag{a}{{\footnotesize open}}
\psfrag{b}{{\footnotesize closed}}
\psfrag{c}{{\footnotesize no big bang}}
\psfrag{z}{{\small $\Omega_M$}}
\psfrag{h}{\small {$\Omega_{r_c}$}}
\resizebox{12cm}{9cm}{\includegraphics{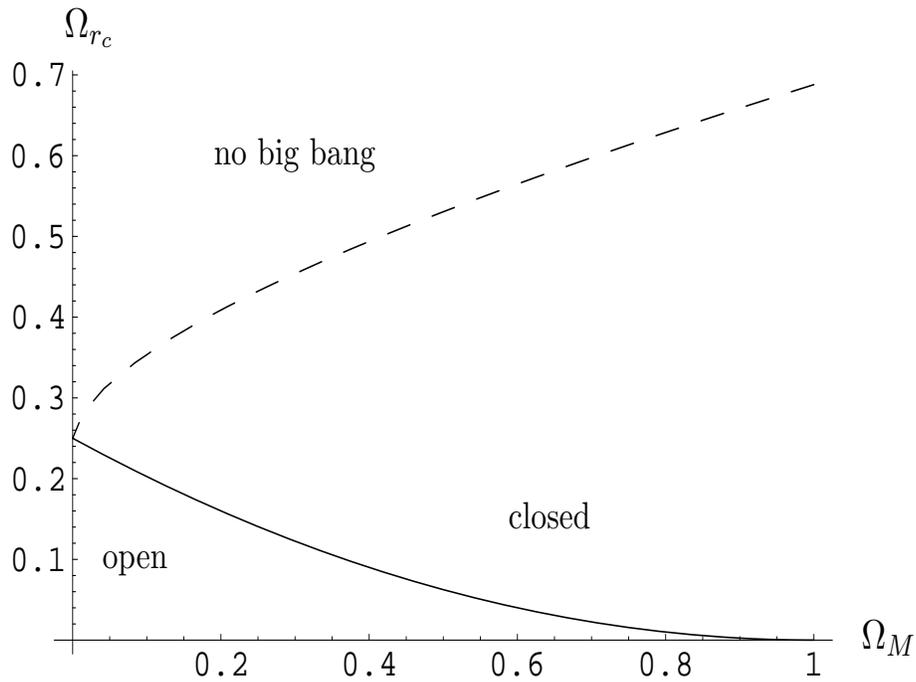}}
\caption{Different possibilities for the expansion 
as a function of $\Omega_M$ and $\Omega_{r_c}$. The solid line denotes a 
flat universe ($k=0$), with  $\Omega_{r_c}$ obtained through equation 
(\ref{flat5}). The Universes above the solid line are closed ($k=1$),
the universes below  are open  ($k=-1$). 
The Universes above the dashed line avoid the big bang singularity by 
bouncing in the past.}
\label{fig0}
\end{figure}

The $z=0$ value of equation  of equation(\ref{HH5D})
leads to the normalization condition: 
\begin{equation}\label{normalize}
 \Omega_k + \left( \sqrt{\Omega_{r_c}} + \sqrt{ \Omega_{r_c} +
  \Omega_M} \right)^2 =1,
\end{equation}
which differs from the conventional relation
\begin{equation}
\Omega_k  + \Omega_M + \Omega_X=1~.
\end{equation}  
For a flat Universe ($\Omega_k=0$) we get from equation 
(\ref{normalize}) 
\begin{equation} \label{flat5}
\Omega_{r_c} = \left(\frac{1-\Omega_M}{2}\right)^2 
\mbox{ and $\Omega_{r_c} <1$}.
\end{equation}
This shows in particular that for a  flat Universe, 
$\Omega_{r_c}$ is always smaller than $\Omega_X$, nevertheless, 
as will be seen below, the effects of  
$\Omega_{r_c}$ and  $\Omega_X$ can be quite similar. 
Figure \ref{fig0} shows the different possibilities 
for the expansion 
as a function of $\Omega_M$ and $\Omega_{r_c}$.

\section{Cosmological Tests}

We would like to discuss now, in a qualitative way, 
a few cosmological tests and measurements. 
We do not expect that the current experimental precision
would enable us to discriminate between the prediction of our  model 
and the ones of standard cosmology. However,  
the future measurements might enable to do so.

In order to  compare the outcome of our model with 
various cosmological tests  we need first to summarize  
some results. In the FLRW metric (\ref{FLRW}), we define, 
as usual (see e.g.\cite{Hogg:1999ad}),  
the transverse, $H_0$ independent (dimensionless),
comoving distance $d_M$: 
\begin{eqnarray}
\begin{array}{llll}
d_M &=& \frac{ S_k \left( \sqrt{|\Omega_k|} d_C \right)}{\sqrt{|\Omega_k|}},
&\mbox{ if $\Omega_k \neq 0$  }, \\
 d_M &=&  d_C ,& \mbox{ if $\Omega_k = 0$ },
\end{array}
\end{eqnarray}
where $d_C$ is defined as follows: 
\begin{equation}
d_C = \int_0^z H_0 \frac{dx}{H(x)}~.
\end{equation}
From the expression for $d_M$ one gets the ($H_0$ independent and
dimensionless) luminosity distance $d_L$ and the 
($H_0$ independent) angular diameter distance $d_A$ given by 
\begin{eqnarray}
d_L &=& (1+z) d_M, \\
d_A &=& \frac{d_M}{1+z}.
\end{eqnarray}
These definitions can be used  on the same footing both 
in standard and in our cosmological  scenarios (as they stand above,
they only rely on the geometry of the four-dimensional Universe
seen by the radiation  which is the same in both cases). 
The only difference is due to the expression for  $H(z)$ 
which enters the definition of $d_C$; one should choose either 
equation (\ref{HH4D}) or (\ref{HH5D}) depending on the case considered.  
Whenever we want to distinguish between the two models, we will put a 
{\it tilde} sign to the quantities corresponding to our model
(e.g. $\tilde{d}_L$).

\subsection{Supernovae Observations}

\begin{figure}
\centering
\psfrag{z}{{\small z}}
\psfrag{h}{\small {$d_L(z)$}}
\resizebox{12cm}{7cm}{\includegraphics{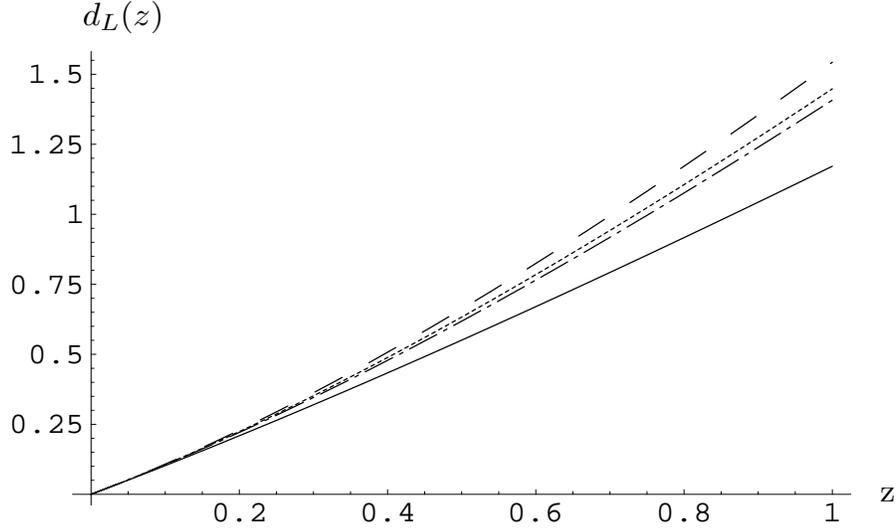}}
\caption{Luminosity distance as a function of red-shift
for ordinary cosmology with $\Omega_\Lambda=0.7, 
\Omega_M=0.3,k=0$ (dashed line),
$\Omega_\Lambda=0, \Omega_M=1,k=0$ (solid line),
and dark energy with $\Omega_X=0.7,w_X=-0.6, \Omega_M=0.3, k=0$ 
(dotted-dashed line) and in our model (dotted line) 
with $\Omega_M = 0.3$ and a flat universe 
(for which one gets from equation (\ref{flat5}) $\Omega_{r_c} = 0.12$
and $r_c =1.4 H_0^{-1}$).}
\label{fig2}
\end{figure}

The evidence for an accelerated universe coming from 
supernovae observation  relies primarily 
on the measurement of the apparent 
magnitude of type Ia supernovae as a function of red-shift.
The apparent magnitude $m$ of a given supernova 
is  a function of its absolute magnitude $\cal M$, 
the Hubble constant $H_0$ and $d_L(z)$ 
(see e.g. \cite{Goliath:2001af}). 
Considering the supernovae as standard candles,  $\cal M$
is the same for all supernovae, so is $H_0$; thus,  
we need only to compare 
$d_L(z)$ in our model with that in standard cosmology.
Figure \ref{fig2} shows the luminosity distance $d_L$
as a function of red-shift in standard cosmology 
(for zero and non-zero cosmological constant) and in our model. 
This shows the expected behavior: 
our model mimics the cosmological constant in producing the late-time 
accelerated expansion. However, as is also apparent from this plot, 
for the same flat spatial geometry and the same amount of 
non-relativistic matter, our model does  not produce exactly the same 
acceleration as a standard cosmological constant,
but it rather mimics the one obtained from a dark energy 
component with $w_X > -1$.

\subsection{Comparison with dark energy}

\begin{figure}
\centering
\psfrag{z}{ {\small z}}
\psfrag{w04}{{\scriptsize $w_X=-0.4$}}
\psfrag{w06}{{\scriptsize $w_X=-0.6$}}
\psfrag{w08}{{\scriptsize $w_X=-0.8$}}
\psfrag{h}{\small {$d_L(z)/d_L^{ref}(z)$}}
\resizebox{12cm}{7cm}{\includegraphics{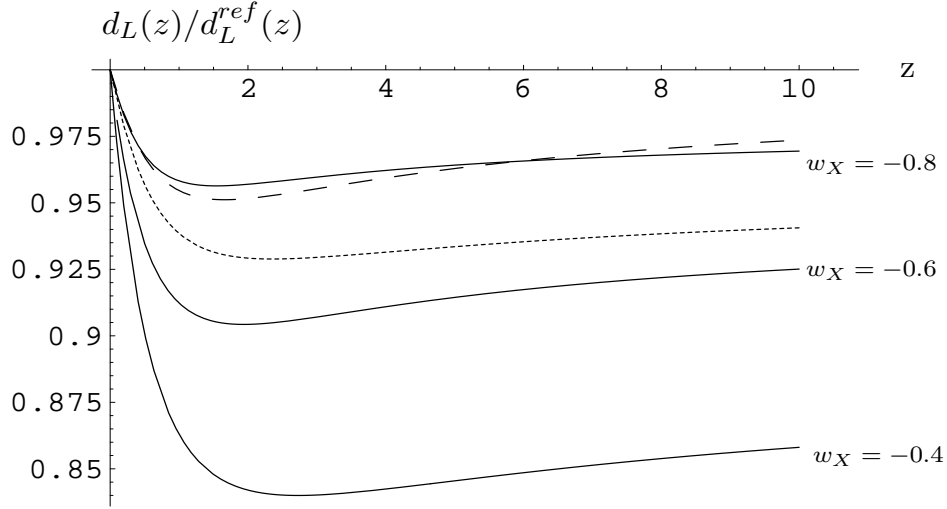}}
\caption{Plot of $d_L(z)/d_L^{ref}(z)$ 
for various models
of dark energy with constant equation of state parameters $w_X$ 
in standard cosmology (solid lines) as compared with the outcome of the  model
consider in this paper (dashed and dotted lines). 
All plots correspond to flat universes with $\Omega_M =0.3$ 
(solid lines, and dotted line), and  $\Omega_M =0.27$ (dashed line).}
\label{fig3}
\end{figure}

\begin{figure}
\centering
\psfrag{z}{ {\small z}}
\psfrag{w04}{{\scriptsize $w_X=-0.4$}}
\psfrag{w06}{{\scriptsize $w_X=-0.6$}}
\psfrag{w08}{{\scriptsize $w_X=-0.8$}}
\psfrag{h}{\small {$d_C(z)H(z)/H^{ref}(z) d_C^{ref}(z)$}}
\resizebox{12cm}{7cm}{\includegraphics{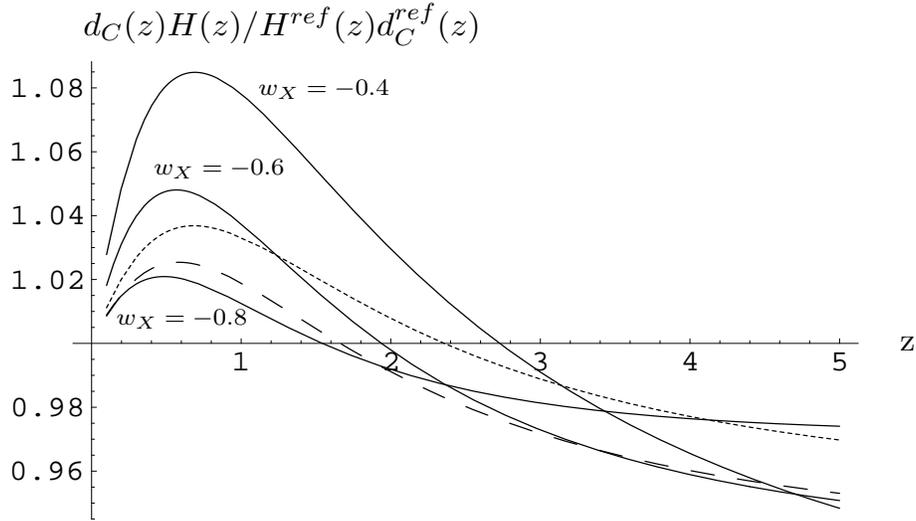}}
\caption{Plot of $H(z) d_C(z)/H^{ref}(z) d_C^{ref}(z)$ (Alcock-Paczynski test)
for various models
of dark energy with constant equation of state parameters $w_X$ 
in standard cosmology (solid lines) as compared with the outcome of the  model
considered in this paper (dashed and dotted lines). 
All plots correspond to flat universes, with $\Omega_M =0.3$ 
(solid lines, and dotted line), and  $\Omega_M =0.27$ (dashed line).}
\label{fig1}
\end{figure}

We want here to compare the predictions of our model to the ones of standard 
cosmology with a dark energy component. For this purpose we choose a reference 
standard model  given by standard cosmology  
with the parameters $\Omega_\Lambda =0.7$, $\Omega_M = 0.3$ and $k=0$ 
(and denote the associated quantities 
 with the superscript $^{ref}$, e.g. $d_L^{ref}$).
Figures \ref{fig3} and \ref{fig1} show respectively the 
luminosity distance $d_L(z)$ and $d_C(z) H(z)$ 
(Alcock-Paczynski test, see e.g.  \cite{Huterer:2000mj}) 
for various cases, showing that with precision tests, one should be able to discriminate between our model and a pure cosmological constant.

\subsection{Cosmic Microwave Background (CMB)}

\begin{figure}
\centering
\psfrag{z}{ {\small $\Omega_M$}}
\psfrag{h}{\small {$\Omega_{r_c}$}}
\resizebox{10cm}{10cm}{\includegraphics{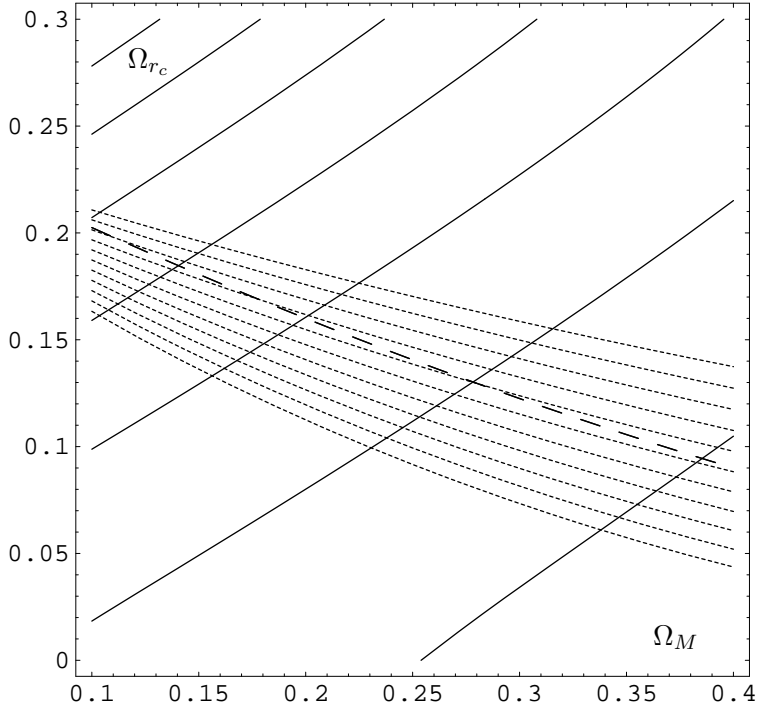}}
\caption{The Solid lines are lines
 of equal luminosity distance (in our model), 
$\tilde{d}_L(z=1)/d^{ref}_L(z=1)$, at red-shift 
$z=1$, the contours are drawn at every 5\% level. The dashed 
line corresponds to a flat universe. The dotted line are line of equal 
$\sqrt{\Omega_M}\tilde{d}_A(z)$ for $z=1100$, the contours are drawn 
at every 5\% level.} 
\label{fig5}
\end{figure}

\begin{figure}
\centering
\psfrag{z}{ {\small $\omega_{X}$}}
\psfrag{h}{\small {$d_A(z=1100)$}}
\resizebox{12cm}{7cm}{\includegraphics{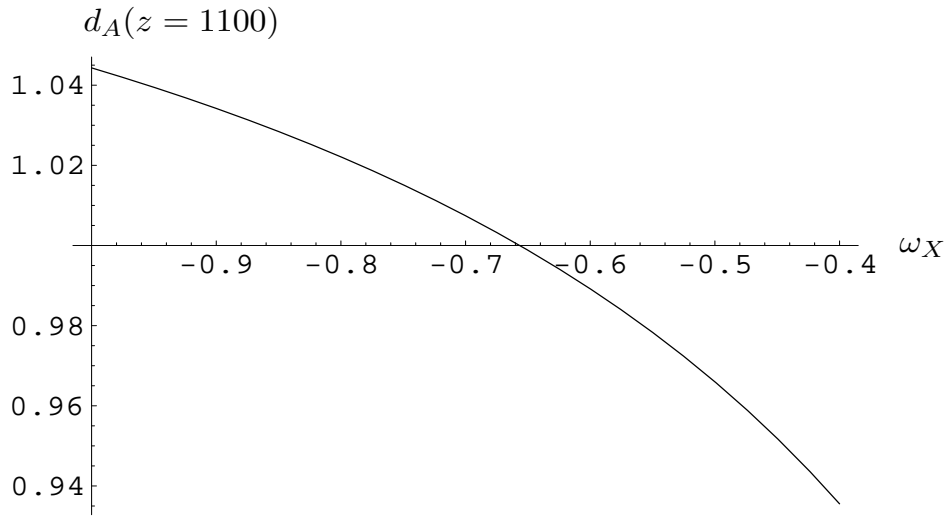}}
\caption{Angular diameter distance $d_A$ at $z=1100$ of standard cosmology
 divided  by $\tilde{d}_A(z=1100)$ in our model, as a function of $w_X$
 for a flat universe and $\Omega_M = 0.3$} 
\label{fig6}
\end{figure}

It is well known that in standard cosmology, the 
location of points of constant 
luminosity distance at small $z$ 
is degenerated in the plane ($\Omega_M$, $\Omega_\Lambda$). This degeneracy 
can be lifted through CMB observations. Figure \ref{fig5} shows that this is 
the case as well in our model (Which should not be too much of a surprise, 
considering the similarities between early cosmology in the two models,
 as well as between the luminosity distances vs red-shift relations). 
The solid lines of figure \ref{fig5} are lines of constant $\tilde{d}_L$ 
at red-shift $z=1$; the dotted lines are lines of constant $\sqrt{\Omega_M} 
d_A$ at red-shift $z=1100$. This latter quantity roughly sets
 the position of the first acoustic peak in the CMB power spectrum,
 since its inverse measures the angular size on the sky of a physical length 
scale at last scattering proportional to $1/\sqrt{\Omega_M}$ 
(as is at first approximation the sound horizon at last scattering). 
Eventually figure \ref{fig6} shows the angular diameter 
distance $d_A$, at $z=1100$,  of standard cosmology divided by 
$\tilde{d}_A$ in our model, 
as a function of $w_X$, for a flat universe and $\Omega_M = 0.3$. 
This shows that, for the same content of matter (and a flat universe), 
the first Doppler peak 
in our model will be slightly on the left of the one obtained in 
standard cosmology with a pure cosmological constant.

\section{Confronting with Unconventional 4D Theories of Gravity}

One might wonder whether it is possible to obtain
the similar cosmological scenario in purely 
four-dimensional theory by introducing additional generally 
covariant  terms in the Einstein-Hilbert action.
The conventional local terms which can be added to 
the 4D theory contain higher derivatives. 
\beq
\mpl^2 ~\sqrt{g}~
\left (R~+~\alpha~{R^2\over \mpl^2}+...\right )~.
\label{hd}
\eeq
Whatever the origin of these terms might be their contributions should 
be suppressed
at distances bigger than  millimeter. 
That  is required by existing precision gravitational measurements. 
This implies  
that at distances of the present Hubble size their contributions are
even more suppressed. 
For instance, from the requirement  
that the contribution of the $R^2$ term 
to the Newtonian interaction  be  sub-dominant 
at distances around a centimeter implies that the relative 
contribution of the $R^2$ term at the Hubble scale is suppressed 
by the factor $({\rm cm}^2 H_0^2)~\sim~10^{-56}$. 
The contributions of other higher terms  
are suppressed even stronger.

It seems that the only way to accommodate this unusual behavior in 
a would-be pure 4D theory of gravity is to introduce terms 
with fractional powers of  the Ricci scalar,
for instance, such as  the term $\sqrt{g~R}$. However, it is hard to make sense of such a theory.

Therefore, we conclude that the scenario 
discussed in the previous sections is 
intrinsically high-dimensional one.

\section{Constraints}

In our framework such a
low five-dimensional Planck scale is compatible with all the 
observations \cite {DGKN}.
In fact, at distances smaller that the present horizon size the 
brane observer effectively sees a single 4D 
graviton which is coupled with the strength 
$1/\mpl$ (instead of a 5D graviton coupled by 
the $1/\M^{3/2}$ strength).

As it was shown in \cite{DGKN}  the high energy
processes place essentially no constraint on the scale $\M$.
This can be understood in two equivalent ways,
either directly in five-dimensional pictures, 
or in terms of the expansion in 4D  modes.

As was shown above, in five-dimensional language the brane 
observer at high energies sees graviton which is 
indistinguishable from the four-dimensional
one;  for short distances 
the propagator of this graviton is that of a 4D theory
\beq
{\tilde G}_R(p, y=0)~\propto~{1\over p^2}~.
\label{G5massless}
\eeq
Moreover, this state couples to matter 
with the $1/\mpl^2$ strength. 
Therefore in all the processes  with typical
momentum $p << 1/r_c$ the graviton production 
must go just like in 4D theory.
For instance, the rate of the graviton production 
in a process with energy $E$ scales  as
\begin{equation}
\Gamma ~\sim ~{E^3 \over \mpl^2}~.
\label{rate}
\end{equation}
The alternative language is that of the mode expansion. From the point of
view of the four-dimensional brane observer a single 5-dimensional
massless graviton is in fact a continuum of four-dimensional states,
with masses labeled by a
parameter $m$
\begin{equation}
G_{\mu\nu}(x,y) = \int dm~ \phi_m(y)~ h^{(m)}_{\mu\nu}(x)~.
\end{equation}
The crucial point is that the wave-functions of the massive modes
are suppressed on the brane as follows
\begin{equation}
|\phi_m(y=0)|^2~ \propto~{1\over 4~+ ~m^2~r_c^2}~.
\end{equation}
This is due to the intrinsic curvature term on the brane
which ``repels'' heavy modes off the brane \cite{DGKN,carena}. 
As a result their production in
high-energy processes on the brane is very difficult.
Let us once again consider bulk graviton production in a
process with  energy $E$ (e.g. star cooling via graviton emission at temperature $T$ of order $E$).
 This rate is given by \cite {DGKN}
\begin{equation}
 \Gamma ~\sim ~{E^3 \over \M^3}~\int_0^{m_{\rm max}} ~dm ~
|\phi_m(0)|^2~.
\end{equation}
Here the integration goes over the continuum of bulk states  up to a maximum
possible mass which  can be produced in a given process $m_{\rm max} \sim E$.
However, since heavier wave-functions are suppressed on the brane by a
factor ${1 \over
m^2r_c^2}$, the integral is effectively cut-off at $m \sim 1/r_c$, which
gives for the rate
\begin{equation}
  \Gamma \sim {E^3 \over \M^3~r_c} \sim {E^3 \over \mpl^2}~.
\end{equation}
This is in agreement with Eq. (\ref{rate}) and in fact
coincides with the rate of production of a single  
four-dimensional graviton, which is
totally negligible. Thus high-energy processes place no constraint on
scale $\M$ \cite {DGKN}. 

Due to the same reason cosmology places no bound on the scale $\M$.
Indeed, the potential danger would come from the fact that the early
Universe may cool via graviton emission in the bulk, which could affect
the expansion rate and cause deviation from an ordinary FLRW cosmology.
However, due to extraordinarily  suppressed graviton emission at high
temperature, the cooling rate due to this process is totally negligible.
Indeed in radiation-dominated era, the cooling rate due to graviton
emission is
\begin{equation}
\Gamma~ \sim~ {T^3\over \mpl^2}~.
\label{rad}
\end{equation}
At any temperature below $\mpl$ this is much smaller that the expansion
rate of the Universe $H ~\sim~ T^2/\mpl$. Thus essentially until $H \sim
\M^3/\mpl^2$ (which only takes place in the present epoch) Universe
evolves as ``normal''.

The only constraint in such a case comes from the 
measurement of Newtonian force, which implies $\M~> ~10^{-3}$eV
(this will be discussed in more detail elsewhere).

\section{Deterioration due to Dissipation}

In the previous sections we established that 
classically the asymptotic form of the 4D metric on the brane is 
that of de Sitter space. 
Here we would like to ask the question whether this
asymptotic form can be modified due to quantum effects. 
This could happen if there is dissipation of the 
energy stored in the expectation value of the 4D Ricci scalar
into other forms which either can radiate into the bulk or be 
red-shifted away on the brane.  
Below we shall identify such a mechanism of potential 
dissipation.

An observer in de Sitter space is 
submerged in a thermal bath with nonzero temperature 
due to Hawking radiation from the de Sitter horizon.
The temperature of this radiation is $T\sim H$.
The crucial point is that 
the energy stored in this radiation can  
dissipate into the bulk in the form of vary long-wavelength 
graviton emission from the brane.
To estimate the rate of this dissipation we can use Eq. (\ref {rad})
with $T\sim H$. The corresponding change of the 
brane energy density in the absence of other forms of 
matter and radiation is given by:
\beq
{ d \rho_{\rm eff} \over  dt}~=~- {H^3\over \mpl^2}~ \rho_{\rm eff}~,
\label{rho}
\eeq 
where $\rho_{\rm eff}~\equiv~\mpl^2~\langle R \rangle $ and 
the Hubble parameter can be written as $H^2~\propto~\langle R \rangle$.
The corresponding decay time is huge $\tau~\sim~10^{137}$ sec.
Therefore, the 4D metric eventually  asymptotes to flat Minkowski space.
Note the crucial difference from the conventional 4D de Sitter space
where the vacuum energy cannot dissipate anywhere due to the Hawking 
radiation.  In our case the existence of infinite volume
bulk is vital.

\section{Infinite Volume and String Theory}

If the recent observations on the cosmological constant are
confirmed  it may be extremely  nontrivial to
describe the accelerated  Universe within
String Theory \cite {Banks,Witten}.
To briefly summarize the concerns
let us consider a generic theory with extra dimensions.
Usually one is looking for a ground state of the theory
with compactified or warped extra dimensions.
In both of these cases there is a length scale which defines
the volume of the extra space.  This scale cannot be bigger
than a millimeter \cite {ADD}.
Therefore, at larger distances 
a conventional 4-dimensional space is recovered.
Astrophysical observations indicate that this latter
asymptotes to the state of 4-dimensional accelerated expansion similar 
 to 4D de Sitter. In which case the following two
 problems may emerge \cite {Banks,Witten}:

\begin{itemize}

\item{An  observer in  dS space sees a finite portion of the space
bounded by event horizon.
In fact, the four-dimensional  dS interval  can be transformed
into the form:
\beq
ds^2_{\rm dS}~=~-\left (1~-~H^2 ~u^2 \right )~d\tau^2~+~
{du^2 \over (1~-~H^2 ~u^2)}~+~u^2~d\Omega_2~.
\label{ds}
\eeq
An observer is always inside of a finite size horizon.
As was argued in \cite {Banks}
physics for any such an observer is described by a finite number
of degrees of freedom\footnote{Indeed, the number 
of degrees of freedom inside
the region bounded by the horizon is finite.
Moreover, physics of the exterior of the horizon can in principle
be  encoded into the information on the horizon.
This latter, according to the Beckenstein-Hawking formula,
has finite entropy and, therefore,
supports a finite number of degrees of freedom.}.
On the other hand, there are an infinite number of 
degrees of freedom in String Theory and it is not obvious
how String Theory can be reconciled 
with this observation.}

\item{Another related difficulty is encountered
when on tries to define
the String Theory S-matrix on dS space.
As we mentioned above,
we  could think of dS space as
a cavity with a shell surrounding it.
This shell has nonzero temperature.
Thus, particles in the cavity are immersed in a thermal bath
and, moreover, there are no asymptotic states of free particles
required for the definition of the S-matrix.
It was shown recently that these problems generically persist
\cite {Sus,Fish} in quintessence models of
the accelerating Universe.}

\end{itemize}

Both of these difficulties  are related to the fact that
in dS space the comoving volume of the region which can be probed in the 
future by an observer is finite (the same discussion applies to  any
accelerating  Universe with $-1<w<-2/3$, where the
equation of state is $p=w\rho$).

The theories with infinite-volume extra dimensions
might evade these difficulties.
The reason is that the accelerating Universe in this case
can be accommodated in a space which is not simply 4-dimensional
dS.  In fact, as we argued in previous sections, although
the space on the brane looks like de Sitter space
for long time, it will asymptote to space with 
no dS horizon in the infinite future.
 
Let us discuss briefly these issues.

We start by  counting the  number of degrees of freedom
which are in contact with a braneworld  observer.
It is certainly true that an observer on the brane is bounded
in the world-volume dimensions
by a  dS horizon. However, there is no horizon in the
transverse to the brane direction. Thus, any observer on a brane
is in gravitational contact with
infinite space in the bulk. In this case,
the infinite number of bulk
modes of higher dimensional graviton participate in 4D interactions
on the brane \cite {DGP,DGKN}.
Therefore, the number of degrees of freedom
needed to describe physics on the brane is infinite.

The problem of definition of the S-matrix might be more subtle.
Below we present a simplest possibility.
The key observation is that
the metric (\ref {nab}) in the bulk is nothing but the
metric of flat Minkowski space.
Indeed, performing the following  coordinate transformation \cite {derul}:
\begin{eqnarray}
Y^0~&=&~A~\left ({r^2\over 4}~+~1~-~{1\over 4 {\dot a}^2}   \right )
-{1\over 2} \int dt ~{ {a}^2 \over {\dot a}^3 }\partial_t
\left({ {\dot a}\over {a} }  \right )~, \nonumber \\
Y^{i}~&=&~A~x^i~, \nonumber \\
 Y^5~&=&~A~\left ({r^2\over 4}~-~1~-~{1\over 4 {\dot a}^2}   \right )
-{1\over 2} \int dt ~{ {a}^2 \over {\dot a}^3 }\partial_t
\left({ {\dot a}\over {a}  }  \right )~,
\end{eqnarray}
where $r^2=\eta_{ij}x^ix^j$ and $\eta_{ij}={\rm diag}(1,1,1)$,
the metric takes the form:
\beq
ds^2~=~-(dY^0)^2~+~(dY^1)^2~+~(dY^2)^2~+~(dY^3)^2~+~(dY^5)^2~.
\label{mink}
\eeq
The brane itself in this coordinate system transforms into the
following boundary conditions:
\beq
-(Y^0)^2~+~(Y^1)^2~+~(Y^2)^2~+~(Y^3)^2~+~(Y^5)^2~=~
{ 1\over H_0^2}~, \nonumber \\
Y^0(t, y=0)~>~Y^5(t, y=0)~.
\label{boundary}
\eeq
Therefore, the space to the right of the brane
is transformed to  Minkowski space with the
boundary conditions (\ref {boundary}). 

On this space the S-matrix could be defined as there are asymptotic
{\it in} and {\it out} states of free particles.
The same procedure can be applied to  the metric on  the left of the
brane. However, the brane space-time being de Sitter, one encounters the same 
 problems to define in and out states for scattering products localized on the brane. This is true as long as one neglects dissipation discussed in section 7, due to which the whole space-time will asymptote to Minkowski space-time for which the mentionned problems do not persist.

Summarizing, the models with infinite-volume extra dimensions
might be a useful ground for describing an accelerating Universe
within String Theory. In addition we point out that
these models  allow to preserve bulk
supersymmetry even if SUSY is broken on the brane
\cite {DGPsusy,Wittencc}.

\vspace{0.4cm}

{\bf Acknowledgments}
\vspace{0.1cm} \\

We would like to thank 
D. Hogg, A. Lue, R. Scoccimarro, A. Vainshtein and M. Zaldarriaga 
for useful discussions. The work of CD  is supported by David and 
Lucille Packard
Foundation Fellowship for Science and Engineering 99-1462 and by NSF Award 
PHY 9803174. The work of GD was supported in part by David and Lucille Packard
Foundation Fellowship for Science and Engineering, by Alfred P. Sloan
foundation fellowship and by NSF grant PHY-0070787. 
The work of GG  is supported by  DOE Grant DE-FG02-94ER408.
GG thanks Physics Departmenf of NYU for hospitality where
this work was done.


\begin{thebibliography}{99}

\bibitem{cc} A.G. Riess et al., {\it Astroph. J} 116, 1009 (1998);\\
S. Perlmutter et al.,
 [astro-ph/9812133];\\
A.~G.~Riess {\it et al.},
[astro-ph/0104455].


\bibitem{CMB1}
 P.~de Bernardis {\it et al.},
 Nature {\bf 404} (2000) 955
 [astro-ph/0004404];\\
 S.~ Hanany {\it et al.} ApJ Lett. {\bf 545} (2000) 5 [astro-ph/0005123].


\bibitem{CMB2}
 C.~Pryke {\it et al.},
 [astro-ph/0104490];\\
 N.~W.~Halverson {\it et al.},
 [astro-ph/0104489];\\
 C.~B.~Netterfield {\it et al.},
 [astro-ph/0104460];\\
 A.~T.~Lee {\it et al.},
 [astro-ph/0104459].




\bibitem{DGP} G.~Dvali, G.~Gabadadze and M.~Porrati,
Phys.\ Lett.\  {\bf B485} (2000) 208
[hep-th/0005016].

\bibitem{DG}G.~Dvali and G.~Gabadadze,
Phys.\ Rev.\ D {\bf 63}, 065007 (2001); [hep-th/0008054].

\bibitem{Cedric}
C.~Deffayet,
Phys.\ Lett.\ B {\bf 502} (2001) 199
[hep-th/0010186].

\bibitem{Csaki} C.~Csaki, J.~Erlich, T.~J.~Hollowood and J.~Terning,
Phys.\ Rev.\ D {\bf 63}, 065019 (2001)
[hep-th/0003076]. 




\bibitem{DGKN}
G.~Dvali, G.~Gabadadze, M.~Kolanovic and F.~Nitti,
[hep-ph/0102216]



\bibitem{Veltman} H. van Dam, M. Veltman, Nucl. Phys.
{\bf B22}, 397 (1970)~.

\bibitem{Zakharov} V.I. Zakharov, JETP Lett. {\bf 12}, 312 (1970)~.

\bibitem{Vainshtein} A.I. Vainshtein, Phys. Lett. {\bf 39B}, 393 (1972)~.

\bibitem{disc} C. Deffayet, G. Dvali, G. Gabadadze, and A. Vainshtein,
in preparation. 

\bibitem{Kogan}
I.~I.~Kogan, S.~Mouslopoulos and A.~Papazoglou,
Phys.\ Lett.\ B {\bf 503}, 173 (2001)
[hep-th/0011138].

\bibitem{Massimo} M.~Porrati,
Phys.\ Lett.\ B {\bf 498}, 92 (2001)
[hep-th/0011152].

\bibitem{Duff} F.~A.~Dilkes, M.~J.~Duff, J.~T.~Liu and H.~Sati,
[hep-th/0102093]\\
M.~J.~Duff, J.~T.~Liu and H.~Sati,
[hep-th/0105008]

\bibitem{bdl}
P.~Bin{\'e}truy, C.~Deffayet and D.~Langlois,
Nucl.\ Phys.\  {\bf B565} (2000) 269
[hep-th/9905012].

\bibitem{bdel}  
P.~Bin{\'e}truy, C.~Deffayet, U.~Ellwanger and D.~Langlois,
Phys.\ Lett.\  {\bf B477}, 285 (2000)
[hep-th/9910219].

\bibitem{Zurab} Z.~Kakushadze,
Phys.\ Lett.\ B {\bf 488}, 402 (2000)
[hep-th/0006059].



\bibitem{Hogg:1999ad}
D.~W.~Hogg,
[astro-ph/9905116]


\bibitem{Goliath:2001af}
M.~Goliath, R.~Amanullah, P.~Astier, A.~Goobar and R.~Pain,
[astro-ph/0104009]

\bibitem{Huterer:2000mj}
D.~Huterer and M.~S.~Turner,
[astro-ph/0012510]

\bibitem{carena}
M.~Carena, A.~Delgado, J.~Lykken, S.~Pokorski, M.~Quiros and C.~E.~Wagner,
[hep-ph/0102172].

\bibitem{Banks} T.~Banks,
[hep-th/0007146]

\bibitem{Witten} E. Witten, ``Quantum Gravity in DeSitter Spaces''
The Strings 2001 Conference; Tata Institute, Mumbai, India,
January 2001; http://www.theory.tifr.res.in/strings/

\bibitem{ADD} N. Arkani-Hamed, S. Dimopoulos, G. Dvali,
Phys.Lett. B429, (1998) 263, and Phys.\ Rev.\  {\bf D59} (1999) 086004
[hep-ph/9807344]~.

\bibitem{Sus} S.~Hellerman, N.~Kaloper and L.~Susskind,
[hep-th/0104180]

\bibitem{Fish} W.~Fischler, A.~Kashani-Poor, R.~McNees and S.~Paban,
[hep-th/0104181].

\bibitem{derul} N. Deruelle and T. Dolezel, 
Phys.\ Rev.\ D {\bf 62} (2000) 103502, [gr-qg/0004021].






\bibitem{DGPsusy} G.~Dvali, G.~Gabadadze and M.~Porrati,
Phys.\ Lett.\  {\bf B484} (2000) 112
[hep-th/0002190].

\bibitem{Wittencc} E.~Witten,
[hep-ph/0002297].

\end{thebibliography}
\end{document}